\documentclass[journal,twocolumn]{IEEEtran}
\usepackage{amsfonts}
\usepackage{times}
\usepackage{graphicx}
\usepackage{latexsym}
\usepackage{dsfont}
\usepackage{amssymb}
\usepackage{amsmath}
\usepackage{cite}
\usepackage{verbatim}
\usepackage{subfigure}

\newcommand{\figref}[1]{{Fig.}~\ref{#1}}

\def\bb0{{\mathbb{0}}}

\def\bb{{\mathbf{b}}}

\def\bp{{\mathbf{p}}}

\def\br{{\mathbf{r}}}

\def\b0{{\mathbf{0}}}

\def\bS{{\mathbf{S}}}

\def\bX{{\mathbf{X}}}

\def\sf0{{\mathsf{0}}}

\usepackage{epstopdf}
\usepackage{enumerate}
\usepackage{algorithmicx}
\usepackage{algorithm}
\usepackage{amsmath}
\usepackage[noend]{algpseudocode}
\usepackage{float}
\usepackage{color}
\usepackage{makeidx}
\usepackage{bbm}
\usepackage{graphicx}
\usepackage{lipsum}
\usepackage{subfigure}
\usepackage{tablefootnote}
\usepackage{multirow}
\usepackage{multicol}
\usepackage{balance}
\usepackage{booktabs}
\usepackage{soul}
\usepackage[normalem]{ulem}
\usepackage{xcolor}

\newcommand{\sref}[1]{{Section}~\ref{#1}}

\newcommand{\comm}[1]{}

\begin{document}

\title{User Identification: A Key Enabler for Multi-User Vision-Aided Communications}
\author{Gouranga Charan and  Ahmed Alkhateeb \thanks{The authors are with the School of Electrical, Computer, and Energy Engineering, Arizona State University. Emails: \{gcharan, alkhateeb\}@asu.edu. This work was supported by the National Science Foundation (NSF) under Grant No. 2048021.} } 

\maketitle

\begin{abstract}
Vision-aided wireless communication is attracting increasing interest and finding new use cases in various wireless communication applications. These vision-aided communication frameworks leverage visual data captured, for example, by cameras installed at the infrastructure or mobile devices to construct some perception about the communication environment through the use of deep learning and advances in computer vision and visual scene understanding. Prior work has investigated various problems such as vision-aided beam, blockage, and hand-off prediction in millimeter wave (mmWave) systems and vision-aided covariance prediction in massive MIMO systems. This prior work, however, has focused on scenarios with a single object (user) in front of the camera. In this paper, we define the \textit{user identification} task as a key enabler for realistic vision-aided communication systems that can operate in crowded scenarios and support multi-user applications. The objective of the user identification task is to identify the target communication user from the other candidate objects (distractors) in the visual scene. We develop machine learning models that process either one frame or a sequence of frames of visual and wireless data to efficiently identify the target user in the visual/communication environment. Using the large-scale multi-modal sense and communication dataset, DeepSense 6G, which is based on real-world measurements, we show that the developed approaches can successfully identify the target users with more than 97$\%$ accuracy in realistic settings. This paves the way for scaling the vision-aided wireless communication applications to real-world scenarios and practical deployments. 
\end{abstract}

\begin{IEEEkeywords}
	Millimeter-wave, user identification, sensing, camera, deep learning, computer vision.
\end{IEEEkeywords}

\section{Introduction} \label{sec:Intro}
The use of millimeter wave (mmWave) and sub-terahertz (sub-THz) bands is essential to meet the demanding data needs of 5G and future technologies \cite{Rappaport2019,Heath2016}. However, these systems rely on the use of large antenna arrays and narrow directive beams at both the transmitter and receiver to guarantee sufficient receive power. Selecting the optimal beams for these large antennas is associated with a large training overhead. This makes it challenging for mmWave/THz communication systems to support highly-mobile wireless applications such as virtual/augmented reality and connected vehicles \cite{andrews2016modeling}. Furthermore, these high-frequency signals are dependent on direct, line-of-sight (LOS) paths to achieve sufficient receive power \cite{bennis2018ultrareliable}. Any obstacles in the environment that block these LOS links can interrupt communication or significantly degrade the link quality. This is primarily due to the high penetration loss of mmWave/sub-terahertz signals, which drastically reduces the received power for non-line-of-sight (NLOS) links \cite{Rappaport2019, andrews2016modeling}.


\begin{figure}[!t]
	\centering
	\includegraphics[width=0.9\linewidth]{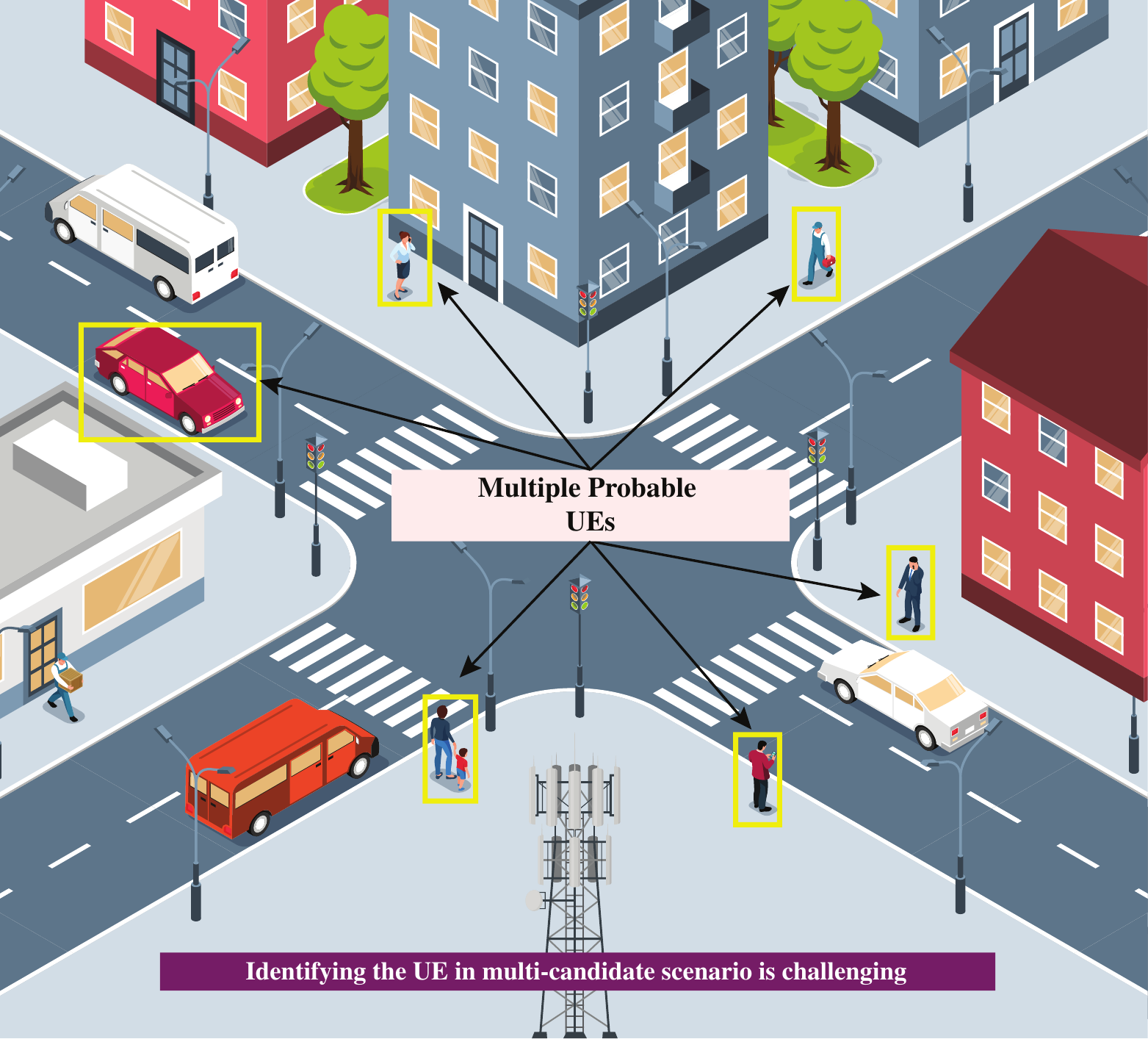}
	\caption{ This figure illustrates the challenges associated with vision-aided communication in multi-candidate scenarios. As shown in the figure, any one or more of the highlighted objects can be the user(s).  The user identification task is the task of identifying which one of the candidate users present in the visual scene in the communication user.}
	\label{fig:UE_identification}
\end{figure}


Leveraging machine learning (ML) to address these challenges has gained increasing interest in the last few years \cite{Alkhateeb2018a,Alkhateeb2018,Lim2021, Alrabeiah2020,Wu2022}. The role of machine learning (and artificial intelligence in general) in tackling problems such as beam training overhead, the sensitivity of mmWave/sub-THz signals to blockages, and demands for low-latency communications has been first investigated using only wireless signals. These solutions, however, are limited in their ability to scale to complex/crowded, or realistic scenarios. This motivated the development of machine learning-based approaches that leverage  side information to overcome the challenges associated with the mmWave/sub-THz communication systems. In order to predict blockages early enough, i.e., before they block the links, solutions based on vision, radar, and LiDAR sensory data were proposed for the first time in \cite{charan2022blockage, demirhan2021radar, wu2021lidar_WCNC, Pinho, Demirhan_mgazine_radar}. Similarly, for fast mmWave/sub-THz beam prediction, solutions based on vision, position,  radar, and LiDAR were proposed in \cite{charan2021c, charan2022drone, morais2022pos, Alrabeiah2020a, Arvinte2019, Wang18, al2022intelligent, demirhan2021beam,Jiang_LiDAR}. These sensing-aided wireless communication solutions were developed, however, for single-candidate scenarios and might not scale to a real-world setting with multiple objects in the environment. Therefore, an important question that arises is \textit{how do we develop sensing-aided wireless communication solutions that can scale to real-world scenarios with multiple objects in the sensing scene?}.

To enable sensing-aided communication systems in real-world settings \cite{alkhateeb2023real}, we need to enable these systems to operate in multi-candidate and multi-user settings.  To illustrate this, consider the example of a sensing-aided beam prediction task. In practice, from the basestation perspective, there can be multiple relevant objects in the wireless environment. Any of those objects can be the object of interest (the user). Therefore, the machine learning models must demonstrate a deep understanding of the wireless environment to be able to predict the optimal beam indices correctly. In particular, it needs to identify the probable user candidate among the different objects in the environment. One main approach to achieve that is by leveraging additional sensing information (attributes) for the user.  This brings the following important question: \textit{How can machine learning models leverage additional sensing data such as position or wireless receive power to identify the target user in the sensing (e.g., visual) scene?} 

In this paper, we focus on visual sensing and attempt to answer this question. The main contributions of the paper can be summarized as follows:
\begin{itemize}
	\item Formulating the user identification problem in vision-aided mmWave/THz wireless communication networks considering practical visual and communication models. 
	\item Developing machine learning approaches that are capable of (i) detecting the objects of interest in the wireless environment and (ii) efficiently identifying the user in the visual scene among the different objects in the environment.
	\item Demonstrating the robustness of our solution in adapting to unseen scenarios, maintaining user identification accuracy of approximately $90\%$, even when trained data from the scenario is absent. This showcases the flexibility of our model, a significant advantage for practical deployment in diverse 5G and beyond base station locations.
	\item Providing the first real-world evaluation of sensing-aided user identification based on our large-scale dataset, DeepSense 6G \cite{DeepSense}, that consists of co-existing multi-modal sensing and wireless communication data. 
\end{itemize}
Based on the adopted real-world dataset, the developed solution achieves $\approx 97\%$ and $\approx 99\%$ user identification accuracy for input sequence lengths of $1$ and $5$, respectively. This highlights the potential of leveraging machine learning and sensing data in addressing the critical task of identifying the user in the scene. In particular, the ability to identify the user in the scene enables the network to make proactive beam/basestation switching decisions and predict future line-of-sight link blockage, enhancing the overall network reliability and latency performance. 

\section{Sensing-Aided User Identification: \\System Model and Problem Formulation}\label{sec:sys_ch_mod}

The utilization of additional sensing data has shown great potential for $5$G and beyond wireless communication systems and can help overcome some of the significant challenges associated with them. However, some fundamental challenges still need to be investigated to develop and implement sensing-aided solutions in the real world. One such real challenge is the ability of the communication system to distinguish between objects transmitting/receiving radio signals in the wireless environment (hereafter referred to as users) and non-transmitting/non-receiving objects (referred to as the distractors). This ability to identify the objects of interest or the users in the wireless environment is referred to as the user identification task. In this work, the user identification task is poised and studied in a mmWave communication setting. In this section, we first present the adopted wireless communication system model in Section~\ref{subsec:system_model} and then formulate the sensing-aided user identification problem in Section~\ref{subsec:prob_form}.

\subsection{System Model}\label{subsec:system_model}

This study considers a realistic communication scenario in which a mmWave basestation is serving a mobile user (vehicle) in a busy environment with various moving objects such as other vehicles and pedestrians, among others. The adopted system model encompasses a basestation with an $M$-element Uniform Linear Array (ULA) and an RGB camera, operating at a mmWave frequency band. This mmWave basestation serves a mobile user (transmitter) that is considered to be equipped with a single antenna for simplicity. The communication system in use employs Orthogonal Frequency-Division Multiplexing (OFDM) transmission with $K$ subcarriers and a cyclic prefix of length D. The basestation utilizes a pre-designed beamforming codebook $\boldsymbol{\mathcal F}=\{\mathbf f_q\}_{q=1}^{Q}$, where $\mathbf{f}_q \in \mathbb C^{M\times 1}$ and $Q$ represents the total number of beamforming vectors. Let $\mathbf h_{k}[t] \in \mathbb C^{M\times 1}$  represent the channel between the mmWave basestation and the mobile user at the $k$th subcarrier and time $t$. Given that the basestation utilizes the beamforming vector $\mathbf{f}_q \in \boldsymbol{\mathcal F}$ to serve the user, the receive signal can be represented as follows:
\begin{equation}\label{eq:sys_mod}
	y_{k}[t] = \mathbf h_{k}^T[t] \mathbf f_q[t]x + n_k[t],
\end{equation}
where $n_k[t]$ is a noise sample drawn from a complex Gaussian distribution $\mathcal N_\mathbb C(0,\sigma^2)$. The transmitted complex symbol $x\in \mathbb C$ need to satisfy the following constraint $\mathbb E\left[ |x|^2 \right] = P$, where $P$ is the average symbol power. The beamforming vector $\mathbf f^{\star}[t] \in \boldsymbol{\mathcal F}$ at each time step t is selected to maximize the average receive SNR and is defined as 
\begin{equation}\label{eq:beam_training}
	\mathbf f^{\star}[t] = \underset{\mathbf f_q[t]\in \mathcal F}{\text{argmax}} \frac{1}{K}\sum_{k=1}^{K} \mathrm{SNR}|\mathbf h_{k}^T[t] \mathbf f_q[t] |^2,
\end{equation}
where $\mathrm{SNR}$ is the transmit signal-to-noise ratio, SNR = $\frac{P}{\sigma^2}$.

\subsection{Problem Formulation} \label{subsec:prob_form}
Given the system model in Section~\ref{subsec:system_model}, we provide the formal definition of the sensing-aided user identification task in this section. For this, a general description of the task is first provided. \textbf{User identification} is a multi-modal machine learning task with the primary objective of \textbf{identifying the user among the different objects present in the wireless environment.} The inputs to the machine learning model are the available sensing and wireless data obtained from the environment. We propose to observe a sequence of RGB images of the wireless environment captured by the camera installed at the basestation and utilize the sensing data along with the mmWave receive power vectors to identify the user in the scene. The wireless channel vector $ \mathbf h$ (as defined in Section~\ref{subsec:system_model}), in general, encodes more detailed information regarding the wireless environment, such as the different propagation paths between the transmitter and the receiver; making it a better alternative for the task as compared to the receive power vector. Nevertheless, in the mmWave communication system, it is challenging to obtain this channel information analytically. 

The user identification task can be formally defined as follow. Let $\bX[t] \in \mathbb{R}^{W \times H \times C} $ denote a single RGB image of the environment captured at the basestation at time instant t, where $W$, $H$, and $C$ are  the width, height, and the number of color channels for the image. Further, let  $\bp[t]$ denote the mmWave receive power vector at the basestation. At any time instant $\tau\in \mathbb Z$, the basestation captures a sequence of RGB images and the mmWave receive power vectors, $\bS[\tau]$, defined as
\begin{equation}
	{\bS}[\tau] = \left\{ \bX[t], \bp[t]  \right\}_{t = \tau-r+1}^{\tau}, 
\end{equation}
where $r \in \mathbb Z$ is the length of the input sequence or the observation window to identify the user.  In particular, at any given time instant $\tau$, the goal in this work is for the basestation to observe the sequence of data samples $\bS[\tau]$ to predict the bounding-box vector $\bb_{\text{Tx}}[{\tau}] \in \mathbb{R}^2$ corresponding to the user in the image samples. In order to identify the user, we define a function $f_{\Theta}$ that {maps} the observed sequence of data samples, $\bS[\tau]$ to a prediction (estimate) of the bounding-box vector, $\hat{\bb}_{\text{Tx}}[{\tau}]$. The function $f_{\Theta}$ can be formally expressed as 
\begin{equation}
	f_{\Theta}: \bS[\tau] \rightarrow \hat{\bb}_{\text{Tx}}[{\tau}].
\end{equation}
In this work, we develop a machine learning model to learn this prediction function $f_{\Theta}$, that takes in the observed sequence of data samples $\bS[\tau]$ and predicts the bounding box of the user $\hat{\bb}_{\text{Tx}}[{\tau}]$. Let $ \mathcal D = \left\lbrace \left (\bS, \bb_{\text{Tx}} \right)_u \right\rbrace_{u=1}^U $ represent the dataset of independent samples consisting of \textit{sensing data-bounding box vector} pairs collected from the real wireless environment, where $U$ is the total number of samples in the dataset. The prediction function is parameterized by $\Theta$  representing the model parameters. The dataset $\mathcal D$ of labeled samples is then utilized to optimize the prediction function $f_{\Theta}$ such that it maintains high fidelity for any samples drawn from this dataset. The optimization function aims to maximize the number of correct predictions over all the samples in the dataset $\mathcal D$. The optimization challenge can be formally stated as
\begin{equation}\label{eq:joint_prob}
	f^\star_{\Theta^\star} = \underset{f_{\Theta}(.)}{\text{argmax}}\prod_{u=1}^{U} \mathbb P(\hat{\bb}_{\text{Tx},u} = \bb_{\text{Tx},u}|\bS_u),	
\end{equation}
where the joint probability distribution in \eqref{eq:joint_prob} is due to the implicit assumption that the samples on $\mathcal D$ are drawn from an independent and identical distribution (i.i.d). 
In the next section, we present our proposed machine learning-based solution for the sensing-aided user identification task.

\section{Sensing-Aided User Identification: \\ A Deep Learning Solution} \label{sec:prop_sol}

In this section, we present an in-depth overview of the proposed sensing-aided user identification solution. First, we present the key idea in Section~\ref{subsec:key_idea} and then explain the details of our proposed solution in Section~\ref{subsec:single_tx_id} and Section~\ref{subsec:sequence_tx_id}

\begin{figure*}[!t]
	\centering
	\includegraphics[width=1.0\linewidth]{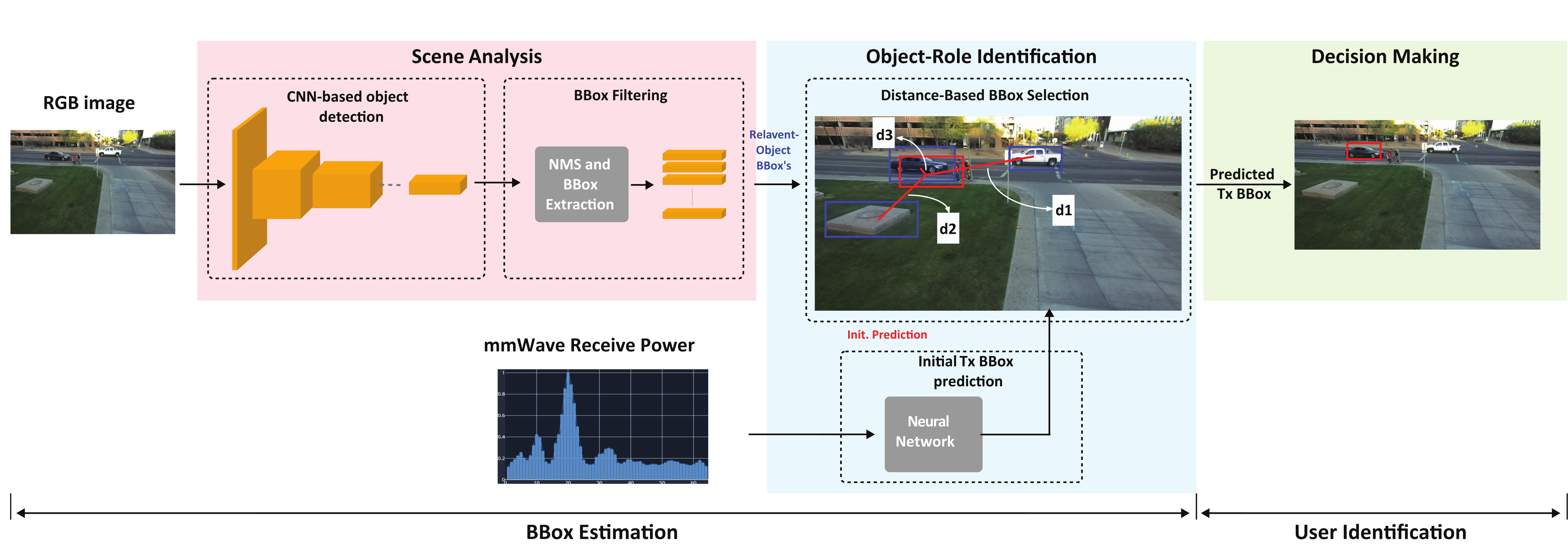}
	\caption{The figure presents the proposed single sample-based user identification model that leverages both visual and wireless data to predict the user in the scene. }
	\label{fig:multi-modal-main-fig}
\end{figure*}

\subsection{Key Idea}\label{subsec:key_idea}

With their large bandwidth, the mmWave/sub-THz communication systems can satisfy the high data rate requirements of several current and future applications. However, communication in these bands is faced with several challenges. One major challenge arises from the high sensitivity of the mmWave/sub-THz signals to blockages. For this, high-frequency signals suffer from significant penetration loss and primarily rely on line-of-sight (LOS) communication. The high-frequency signals, further, suffer from severe path loss. To overcome this huge path loss, the mmWave/sub-THz communication systems must deploy large antenna arrays and use narrow directed beams to guarantee a sufficient receive signal-to-noise ratio (SNR). \textbf{This dependence of the mmWave/sub-THz systems on LOS links and the usage of directive radiation patterns form the basic building block of our proposed user identification solution.} 

The directivity of antenna arrays can be visualized as a way of concentrating the emitted radiation in a single direction. For ULAs, this directivity is achieved by the beamforming vectors in the pre-defined codebook $\boldsymbol{\mathcal F}$. The beamforming vectors can be envisioned as slicing the scene (spatial dimension) into multiple (possibly overlapping) sectors, where each sector is associated with a particular beam value. This sectoring of the wireless environment by the beamforming vectors can be extended to a visual scene. Note that the RGB image is merely a projection of the $3$D space onto a $2$D image plane. The sectoring induced by the beamforming vectors can then be projected onto the $2$D image plane, resulting in the form of image sectoring. Therefore, the knowledge of the optimal beamforming vector or the receive power vector, in general, can be translated to directional information in an image, i.e., the direction from which the current received signal arrived.

Furthermore, the recent advancements in machine learning and computer vision have enabled several new capabilities, such as object detection, multi-object tracking, and image segmentation, to name a few. Therefore, utilizing state-of-the-art object detection models makes it possible to identify different objects in the wireless environment with high fidelity in near real-time. The fast and accurate object detection capabilities paired with the directional information obtained from the receive power vectors can enable us to differentiate between the objects of interest (user) from the distractors in the scene. 

\begin{figure*}[!t]
	\centering
	\includegraphics[width=1\linewidth]{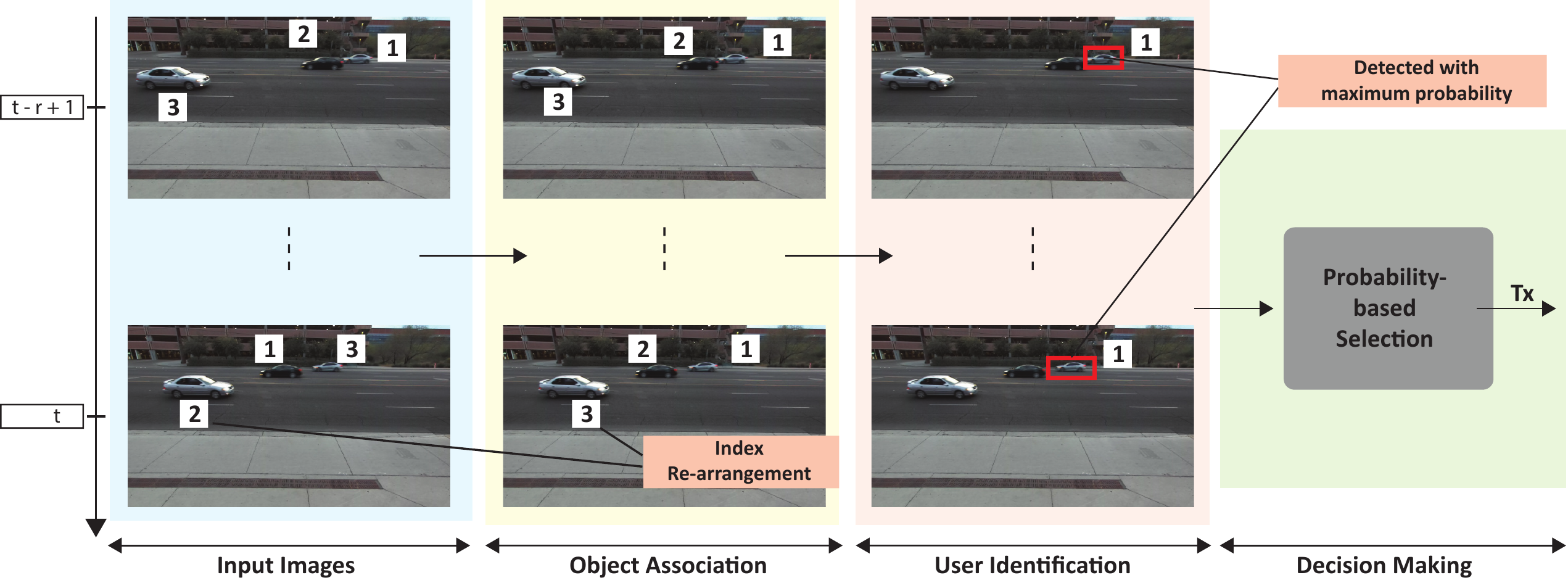}
	\caption{The figure presents the sequence-based user identification model that leverages both visual and wireless data to predict the user in the scene. It shows the three proposed steps: (i) object association-based tracking, (ii) user identification, and (iii) maximum probability-based identification.}
	\label{fig:sequence_tx_id}	
\end{figure*}

Based on the idea proposed above, theoretically, detecting the users with just one data sample is possible. The question that arises now \textbf{is one data sample enough for correctly identifying the user?} In order to answer this question, we first need to understand the challenges associated with this approach. There are primarily three main challenges: (i) Object detection models are not perfect. There is a possibility that several objects, including the users, might not be detected, which might result in false detection. (ii) The user can be partially or entirely occluded in a particular instance. Relying on just that one sample to identify the user will result in a wrong prediction. (iii) One of this solution's key components is identifying the user's approximate location in the scene by utilizing the optimal beamforming or receive power vectors. It is essential to point out here that generating sharp and directive beams with no side lobes are challenging due to the non-idealities and impairments in the hardware. Such hardware limitations will result in non-ideal sectoring and lead to a distractor being labeled as a user. One possible solution to overcome these challenges is to observe a sequence of data samples (image and wireless data) to determine the user accurately. At any given instant $\tau$, by observing a sequence of $r$ current and previous samples, we inherently reduce the effect of non-ideal sectoring and the probability of missed detection. Therefore, in conclusion, although it might be possible to detect the user using just one pair of image-wireless data, it is imperative to observe a sequence of data samples to increase the probability of correctly identifying the user in the environment. In this work, we propose to utilize a sequence of $r$ image and wireless data samples to predict the user in the scene with high fidelity.

\subsection{A Single Sample-based Approach}\label{subsec:single_tx_id}

The following subsection outlines the proposed solution for identifying users within a multi-candidate, real-world wireless setting. We propose a novel approach that leverages visual and wireless data from the dataset $\mathcal D$ to identify the user in the scene accurately. We will first present the solution for identifying the user using one data sample and then extend the proposed solution for a sequence-based approach. A three-step architecture is proposed for the single data sample-based user identification task. The first stage of this framework involves the use of Deep Neural Networks (DNNs) to generate bounding boxes that encapsulate different objects present in the scene. This step is performed to identify and locate all relevant objects within the environment. In the second stage, we propose a DNN that utilizes wireless data to predict the likely centers of the user's bounding boxes. The final stage consists of filtering out detected candidates that are not the radio transmitter/receiver. A comprehensive explanation of the three-step DNN structure is presented in the following paragraphs. A detailed depiction of the user identification solution based on a single data sample is presented in Figure \figref{fig:multi-modal-main-fig}.

\textbf{(i) Bounding box detection:} In order to perform user identification in real wireless environments, the first step involves identifying all the relevant objects of interest within the scene, a process termed ``scene analysis.'' For example, in a scene depicting a city street, relevant objects include but are not limited to, cars, trucks, buses, pedestrians, and cyclists. A pre-trained state-of-the-art object detector is adopted for this purpose. We utilize a COCO \cite{lin2014microsoft} pre-trained YOLOv3 \cite{Yolo, yolov3} architecture for the task of bounding box detection in this research due to its ability to deliver accurate detection at a relatively high frame rate, thereby reducing inference latency. However, to train and assess the proposed user identification solution, it is critical to have the precise bounding-box coordinates (ground truth) of the user within the scene. While a pre-trained object detection model like YOLO can accurately identify the relevant objects, it cannot deliver detailed information specifying which object is the user. Thus, it necessitates the process of manual annotation and fine-tuning of the pre-trained object detection model. For this, the object detection model is further fine-tuned to identify two classes of objects within the scene, labeled as ``User'' and ``Distractor''. To fine-tune the object detection model, we manually annotate a subset of the dataset as mentioned in \sref{sec:dev_data}. Specifically, a portion of the dataset (images) is manually annotated to label the relevant objects, where the radio transmitter/receiver is tagged as ``User'' and all other relevant objects as ``Distractor''. The modified object detector is then fine-tuned in a supervised manner using this labeled dataset. The refined object detection model is then utilized to generate the bounding-box coordinates of the remaining samples in the dataset. In order to ensure the accuracy of the generated bounding boxes, they further undergo a manual verification process. During inference, the fine-tuned YOLOv3 model generates bounding boxes for the detected candidates in the scene and their confidence scores. The output bounding boxes are then utilized to construct the relevant-object matrix $\textbf{B} \in \mathbb{R}^{N \times 2}$ such that each row contains only the normalized coordinates of the center of a bounding box, with $N$ representing the number of relevant objects in the scene.

\textbf{(ii) Bounding box center prediction:} This step involves utilizing both the relevant-object matrix $\mathbf B$ and the wireless received power vector to predict the bounding box center coordinates of the user. It includes learning a prediction function that estimates the bounding box center coordinates of the user using the receive power vector. The primary goal is to encode the relationship between the received power vector and the object's location in the image. This function is learned using a 2-layered feed-forward neural network 
	\begin{equation}
		f_{\Theta2}: \br[t] \rightarrow \hat{\bb}_{\text{Tx}}[{t}],
	\end{equation}
where $\hat{\bb}_{\text{Tx}}[{t}] \in \mathbb R^{2 \times 1}$ is a vector with an initial prediction of the centers of the user and the $\br[t] \in \mathbb R^{Q \times 1}$ is the mmWave receive power vector at any time instant $t$. Let $\mathcal D_2 = \left\lbrace \left (\br, \bb_{\text{Tx}} \right)_u \right\rbrace_{u=1}^U$ and $\mathcal D_2  \subset \mathcal D$ be a dataset comprising of the mmWave receive power vectors and the ground-truth bounding box center coordinates of the user. The prediction function $f_{\Theta2}$ is parameterized by a set $\Theta_2$, which represents the model parameters and is learned from the dataset $\mathcal D_2$ of the labeled data samples. As $\hat{\mathbf b}_{\text{Tx}}$ is an initial estimate relying solely on receive power vector, it is not expected to be the final prediction but merely an approximation. The idea here is to utilize this initial approximate prediction in conjunction with the relevant-object matrix $\mathbf B$ to identify (or select) the object that is the source of the radio signal.

\textbf{(iii) Bounding box selection:} The previous two steps, i.e., bounding box detection and the bounding box center prediction, provide two vital pieces of information: (i) The first stage helps identify all the objects in the environment. More specifically, it outputs the relevant-object matrix $\mathbf B$ comprising the bounding box coordinates of all the objects of interest (probable users) in the wireless environment. (ii) In the second step, we leverage the additional modality, i.e., the wireless receive power vector, to predict the approximate center coordinates of the user in the scene. We aim to utilize these two pieces of information to identify the user within the scene accurately. The final identification process is performed using the nearest neighbor algorithm with a Euclidean distance metric. We first compute the Euclidean distance between the predicted center coordinates and all objects in $\mathbf B$. The object in $\mathbf B$ with the shortest distance to $\hat{\mathbf b}_{\text{Tx}}$ is selected \textit{as the nearest neighbor, consequently, identified as the predicted user object.} The underlying assumption is that a well-tuned prediction function $f_{\Theta2}$ can approximate the center coordinates closely to the actual values, thereby facilitating accurate user detection via the Euclidean distance-based metric.


\begin{figure*}[!t]
	\centering
	\includegraphics[width=1\linewidth]{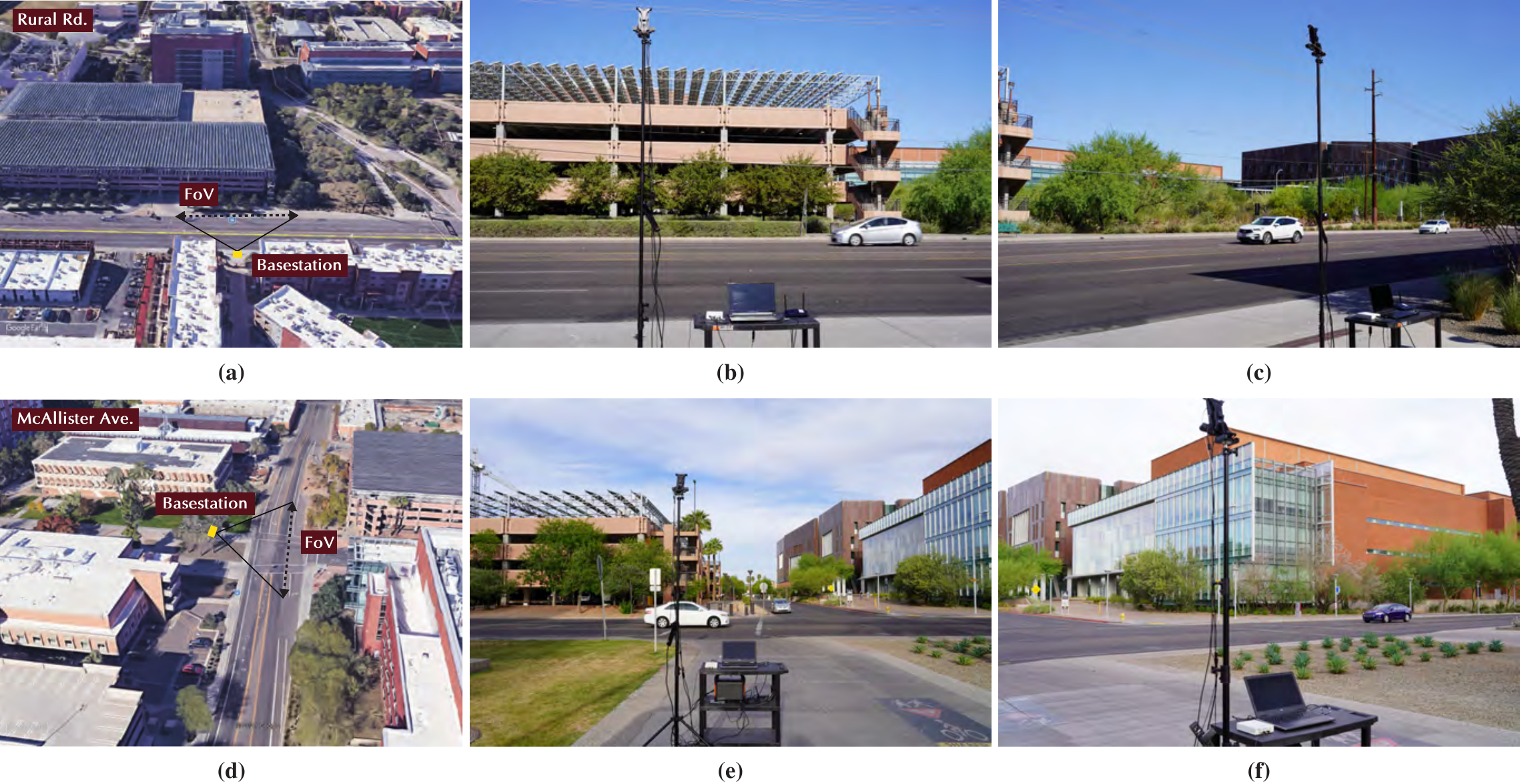}
	\caption{This figure presents the overview of the DeepSense 6G testbed and the location used in scenarios $1$, $3$ and $4$. Figures (a) shows the Google map top-view of the data collection location for scenarios $3$ and $4$. In figures (b) and (c), we show the DeepSense $6$G testbed deployed in these locations. Figures (d), (e), and (f) show the Google map top-view and the testbed deployment for scenario $1$.}
	\label{fig:deepsense}
\end{figure*}


\subsection{A Sequence-based Approach}\label{subsec:sequence_tx_id}

The three-step solution proposed in \sref{subsec:single_tx_id} can help identify the user from one data sample. The question that now arises is \textbf{how do we extend this solution to determine the probable users from a sequence of data samples?} It is essential to note here that, in this work, we do not consider the no-user scenario; in other words, data collected at every time step $t$ will have a user present in the wireless environment. The underlying principle of the sequence-based user identification task is as follows: \textbf{Instead of relying on just one data sample to identify the user, the proposed approach observes a sequence of $r$ data samples. The object identified as a user, the maximum number of times in these $r$ consecutive samples is tagged as the user.} However, moving from a single-sample-based solution to a sequence-based approach has its own challenges. \textbf{This primarily arises from the difficulty in ensuring that the objects detected as users in two consecutive time steps are the same.} Given a sequence of $r$ image samples, the user should be present in every image. Furthermore, as mentioned in \sref{subsec:system_model}, we consider a mobile user in this work. Therefore, the location of the user and other objects in the wireless environment and the RGB image is not fixed across these $r$ consecutive samples. Therefore, to truly perform sequence-based user identification, we also need to track all the relevant objects through time, in addition to identifying which of these objects is the user in the scene. The proposed sequence-based user identification solution comprises three steps: (i) Object association-based tracking, (ii) user identification, and (iii) maximum probability-based identification. 

\textbf{(i) Object association-based tracking:} The field of Multiple Object Tracking (MOT) \cite{wang2020towards, meinhardt2022trackformer, luo2021multiple} has been a focal point of active research, and several state-of-the-art algorithms have been proposed. However, our study diverges from traditional approaches in that it focuses on vehicle-to-infrastructure communication, with mobile vehicles serving as the primary objects of interest. While cutting-edge object detection models can effectively detect different objects and yield bounding box coordinates, they do not provide the object IDs. In the context of our problem statement, which employs a sequence of images, the need for object detection is not sufficient on its own. It is equally crucial to assign a unique ID to each detected object and to maintain that ID for as long as the object remains visible in the sequence of images. As such, we have developed a simple, distance-based tracking algorithm instead of using advanced MOT algorithms. In particular, we adopt a Euclidean distance-based measurement technique, similar to the bounding box selection step described in \sref{subsec:single_tx_id}. The first stage of the proposed object association-based tracking algorithm is to detect the different objects of interest across the $r$ image samples in the sequence and extract the bounding box center coordinates. Let us assume that there are $N1$ and $N2$ detected objects in the first and second image of the sequence with different objects labeled from $1, \ldots, N1$ for the first image and labeled $1, \ldots, N2$ for the second image. Now there are two possibilities: (i) The same number of objects in two consecutive image samples, i.e., $N1=N2$, and (ii) a different number of objects, i.e., either $N1 > N2$ or $N1 < N2$. In the second stage of the algorithm, we calculate the Euclidean distance between each detected object in the first image and the objects in the second image. The objects in the second image are then re-numbered based on this calculated distance. For example, if the $3$rd object in the first image has the shortest Euclidean distance with the $1$st detected object in the second image, then this object is re-numbered as $3$. The principle behind this algorithm is that for two consecutive image samples, the distance between the bounding box center coordinates will be the least for the same object compared to other objects in the scene. Therefore, this algorithm helps identify and track the different objects across the $r$ image samples.

\begin{figure*}[!t]
	\centering
	\includegraphics[width=1\linewidth]{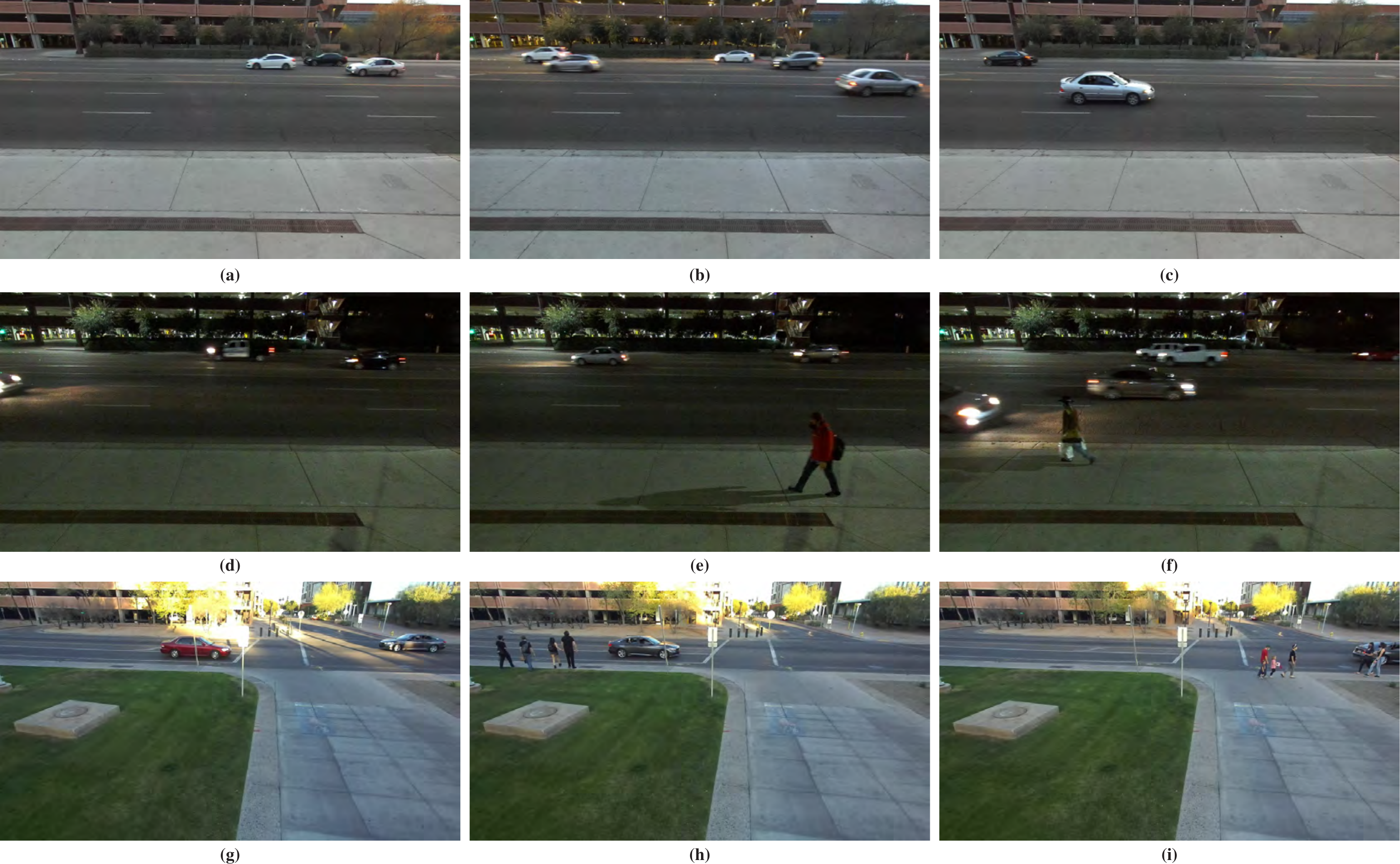}
	\caption{In this figure, we present the different image samples of scenarios $1$, $3$, and $5$. Figures (a), (b), and (c) are from scenarios $3$. Images in (d), (e), and (f) are taken from scenario $4$. It shows the different lighting conditions (day, dusk, and night) in which the dataset was collected, highlighting the diversity in these scenarios. In figures (g), (h), and (i), we present the image samples from scenario $1$. The contrasting elements between scenarios $3$, $4$, and $1$ are distinctly highlighted in these images.}
	\label{fig:deepsense_scenario_photos}
\end{figure*}


\textbf{(ii) User identification:} This step is similar to the single-sample-based user identification described in \sref{subsec:single_tx_id} and is performed for all the $r$ data samples in the sequence. In combination with step one, i.e., object association-based tracking, for each data sample in the sequence, the output of this step is $\hat N \in \mathbb R^{1}$, denoting the index of the user in the scene.

\textbf{(iii) Maximum probability-based identification} The final step of the proposed sequence-based solution is to detect the user based on the observed sequence data accurately. The input to this step is the vector of $r$ indices, i.e., $\{\hat{N}_1, \ldots, \hat{N}_r \}$ that are obtained by performing object association-based tracking and user identification in steps one and two, respectively. The object that has been identified as the user most often across the $r$ data samples is finally identified as the user in the scene.

\section{Testbed Description and Development Dataset}\label{sec:datset}

In this work, we utilize the \textbf{DeepSense 6G} \cite{DeepSense} dataset to evaluate the proposed sensing-assisted user identification solution. DeepSense 6G is a real-world multi-modal dataset designed to facilitate the development of sensing-aided wireless communication applications. The DeepSense 6G dataset consists of co-existing multi-modal data, including vision, mmWave wireless communication, GPS data, LiDAR, and radar, all collected in a real-wireless environment. This section first provides a brief overview of the DeepSense 6G scenarios adopted in this work. Next, we present the final development dataset that has been utilized for the evaluation of the proposed solution.


\begin{table}[!t]
	\caption{Number of Data Sequences in the Development Dataset}
	\centering
	\setlength{\tabcolsep}{5pt}
	\renewcommand{\arraystretch}{1.5}
	\begin{tabular}{|c|cccc|}
		\hline
		\multirow{3}{*}{\textbf{Number of Objects}} & \multicolumn{4}{c|}{\textbf{Number of Sequences}}                                                                                \\ \cline{2-5} 
		& \multicolumn{2}{c|}{\textbf{Scenario 3}}                                  & \multicolumn{2}{c|}{\textbf{Scenario 4}}             \\ \cline{2-5} 
		& \multicolumn{1}{c|}{\textbf{Training}} & \multicolumn{1}{c|}{\textbf{Validation}} & \multicolumn{1}{c|}{\textbf{Training}} & \textbf{Validation} \\ \hline \hline
		1                                           & \multicolumn{1}{c|}{376}            & \multicolumn{1}{c|}{140}            & \multicolumn{1}{c|}{417}            & 187            \\ \hline
		2                                           & \multicolumn{1}{c|}{291}            & \multicolumn{1}{c|}{86}             & \multicolumn{1}{c|}{325}            & 140            \\ \hline
		3                                           & \multicolumn{1}{c|}{140}            & \multicolumn{1}{c|}{46}             & \multicolumn{1}{c|}{182}            & 79             \\ \hline
		4                                           & \multicolumn{1}{c|}{61}             & \multicolumn{1}{c|}{28}             & \multicolumn{1}{c|}{83}             & 30             \\ \hline
		5                                           & \multicolumn{1}{c|}{32}             & \multicolumn{1}{c|}{12}             & \multicolumn{1}{c|}{27}             & 12             \\ \hline
		6                                           & \multicolumn{1}{c|}{6}              & \multicolumn{1}{c|}{7}              & \multicolumn{1}{c|}{6}              & 9              \\ \hline
		7                                           & \multicolumn{1}{c|}{0}              & \multicolumn{1}{c|}{3}              & \multicolumn{1}{c|}{7}              & 0              \\ \hline 
	\end{tabular}
	\label{ref:tab_dataset_analysis}
\end{table}


\subsection{DeepSense 6G: Testbed 1}  \label{sec:testbed}

In this work, we adopt multiple scenarios (1, 3, and 4) from the DeepSense 6G dataset that are specifically designed to explore high-frequency wireless communication applications in a multi-candidate setting. The DeepSense 6G testbed and data collection locations are illustrated in Fig.~\ref{fig:deepsense}. In order to collect data for these scenarios, we utilize the DeepSense testbed $1$, which comprises: (i) a stationary unit (serving as the base station) and (ii) a mobile transmitter (a vehicle). The stationary unit {unit1 (RX)} is equipped with a standard-resolution RGB camera and a mmWave Phased array. This unit deploys a $16$-element ($M = 16$) phased array operating in the $60$ GHz-band and receives the transmitted signal utilizing an over-sampled codebook of $64$ pre-defined beams ($Q = 64$). The mmWave phased array and the RGB camera are positioned such that their fields of view align. As for the mobile unit, {unit2 (TX)}, it is a vehicle equipped with a quasi-omni antenna, constantly transmitting (omni-directional) in the $60$ GHz band and a GPS antenna/receiver to collect the real-time position of the user. Data is captured at a frequency of $\approx 10$ Hz on the basestation side. Each collected data sample comprises an RGB image of the wireless environment, and a $64$-element mmWave receive power vector. For more detailed information regarding the data collection setup and testbed, please refer to \cite{DeepSense}.

\subsection{DeepSense 6G: AI-Ready Dataset} \label{sec:dev_data} 

This work utilizes scenarios $1$, $3$, and $4$ of the DeepSense $6$G dataset. The adopted DeepSense scenarios include diverse data collected at different locations and during different times of the day (day and night). In particular, scenarios $3$ and $4$ are collected at the same location (Rural Rd., Tempe) but at different times of the day. Scenario 1 is collected at a different location (McAllister Ave., Tempe) and primarily consists of data collected during the day time. In Figure~\ref{fig:deepsense_scenario_photos}, we present sample dataset images from scenarios $1$, $3$, and $4$, which highlights the diversity in these scenarios. At any given time instant, $t$, the multi-modal scenario dataset comprises the following: An RGB image, $\bX[t]$, the corresponding receive power vector $\br [t]$ and the user position. We further generate the ground-truth bounding box center coordinates of the user (transmitter) in the scene, $\bb_{\text{TX}}[t]$ (manually labeled). To form the development dataset of the user identification prediction task described in \sref{subsec:prob_form}, the offered DeepSense data is further processed using a sliding window to generate a time-series dataset consisting of $1, 3, \text{and}, 5$ input data images ($r = 1, 3, \text{and},  5$) and the corresponding mmWave receive power. The final step in the processing pipeline is dividing the dataset into training and test sets following a $70-30\%$ split. In Table~\ref{ref:tab_dataset_analysis}, we present the details of the development datasets for the sensing-aided user identification task. 

The main objective of this work is to develop a multi-modal user-identification solution. To evaluate the efficacy of the proposed sensing-aided user-identification solution, we first utilize the development datasets of scenarios $3$ and $4$. The proposed solution is trained and tested on the development dataset of scenarios $3$ and $4$. Given that these scenarios are collected at the same location, but at different time of the day, it also helps to investigate the proposed solution's ability to generalize across different input data distribution. For this, we train the model with the labeled dataset of one of the scenarios and test on the dataset of the other scenario. Next, to analyze the model's ability to adapt to unseen dataset, we utilize the scenario 1 development dataset. It involves training the proposed machine learning-based model on the development dataset of scenario $3$ or $4$ and testing on scenario $1$ dataset.


\begin{table}[!t]
	\caption{Design and Training Hyper-parameters}
	\centering
	\setlength{\tabcolsep}{5pt}
	\renewcommand{\arraystretch}{1.5}
	\begin{tabular}{@{}l|c@{}}
		\toprule
		\toprule
		\textbf{Parameters}                      & \textbf{MLP}        \\ \midrule \midrule
		\textbf{Batch Size}                      & 32                  \\
		\textbf{Learning Rate}                   & $1 \times 10 ^{-3}$ \\
		\textbf{Learning Rate Decay}             & epochs 80 and 120    \\
		\textbf{Learning Rate Reduction Factor}  & 0.1                 \\
		\textbf{Dropout}                         & 0.3                 \\
		\textbf{Total Training Epochs}           & 150                  \\ \bottomrule \bottomrule
	\end{tabular}
	\label{tab:design_param}
\end{table}

\section{Experimental Setup:} \label{sec:exp_setup}
In this section, we will delve into the specifics of the neural network training parameters and the evaluation metrics used. As described in Section~\ref{sec:prop_sol}, the proposed single sample-based user identification solution comprises three steps. As part of the second step, i.e., bounding box center prediction, the mmWave receive power vectors are provided as input to the feed-forward neural network to predict the approximate bounding box center coordinates of the user. The two-layered feed-forward neural network is trained using the labeled development dataset discussed in \sref{sec:dev_data}, employing a cross-entropy loss function and the Adam \cite{Adam} optimizer. All simulations were conducted on a single NVIDIA Quadro 6000 GPU leveraging the PyTorch deep learning framework. The specific design and training hyper-parameters are outlined in Table~\ref{tab:design_param}. The primary method of evaluating the proposed solution is through the top-1 accuracy metric. The definition of top-1 accuracy is as follows:
\begin{equation}
	Acc_{top-1} = \frac{1}{U} \sum_{u=1}^{U} \mathbbm{1} \{\hat{\bb}_{\text{TX}, u}[\tau] = \bb_{\text{TX},u}[\tau] \},
\end{equation}
where $\hat{\bb}_{\text{TX}, u}[\tau]$ and $\bb_{\text{TX},u}[\tau]$ are the predicted and ground-truth bounding box center coordinates, respectively. $U$ is the total number of samples present in the validation/test set. $\mathbbm{1}\{.\}$ is the indicator function.

\section{Performance Evaluation} \label{sec:perf_eval}
This section presents the detailed evaluation of the proposed sensing-aided user identification solution. 


\begin{figure}[!t]
	\centering
	\includegraphics[width=1.0\linewidth]{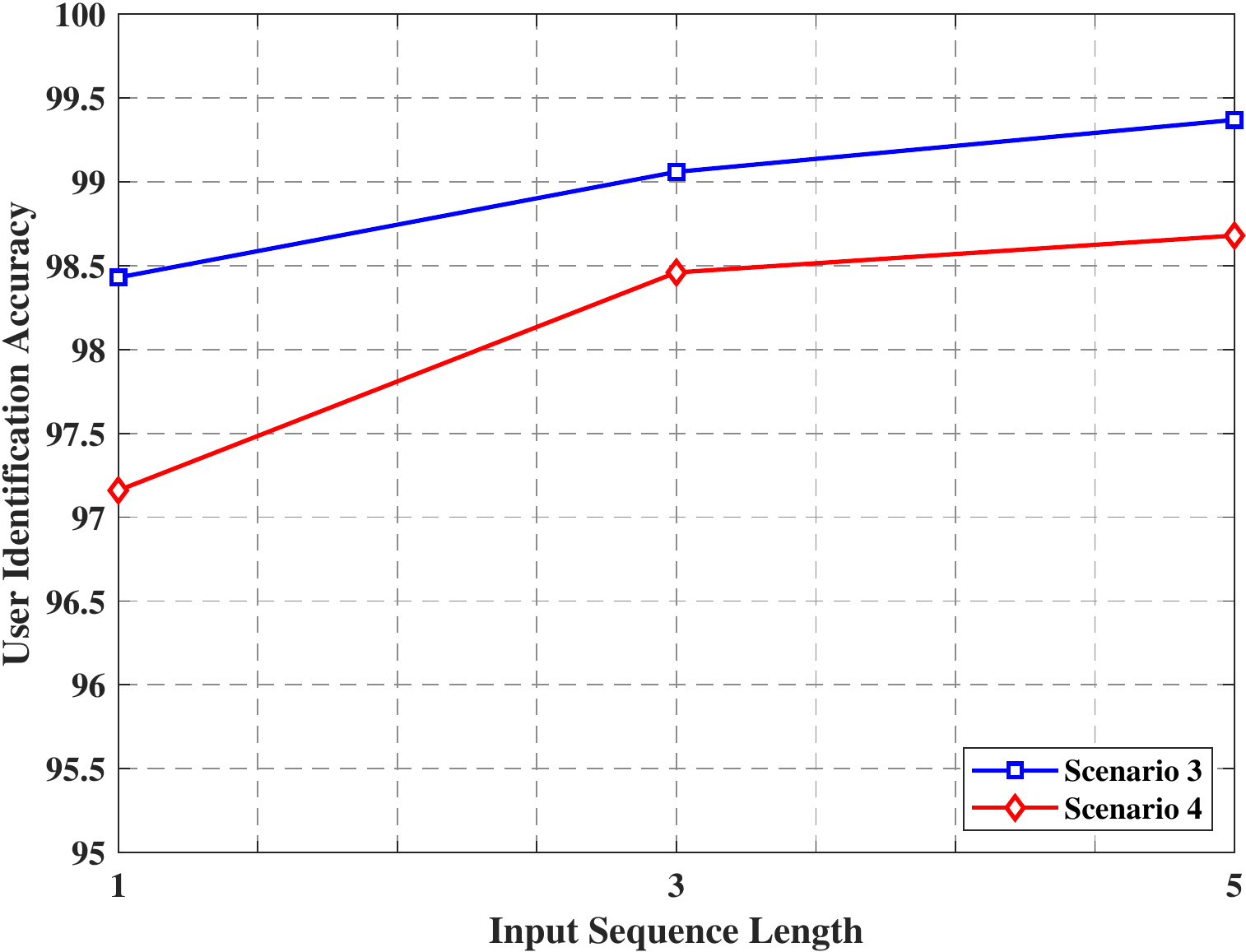}
	\caption{This figure presents the user identification accuracy versus the input sequence length of 1,3 and 5 for both scenarios 3 and 4. It is observed that observing a sequence of samples help in improving the identification accuracy. }
	\label{fig:combined_scene_score}
\end{figure}

\begin{figure}[!t]
	\centering
	\includegraphics[width=1.0\linewidth]{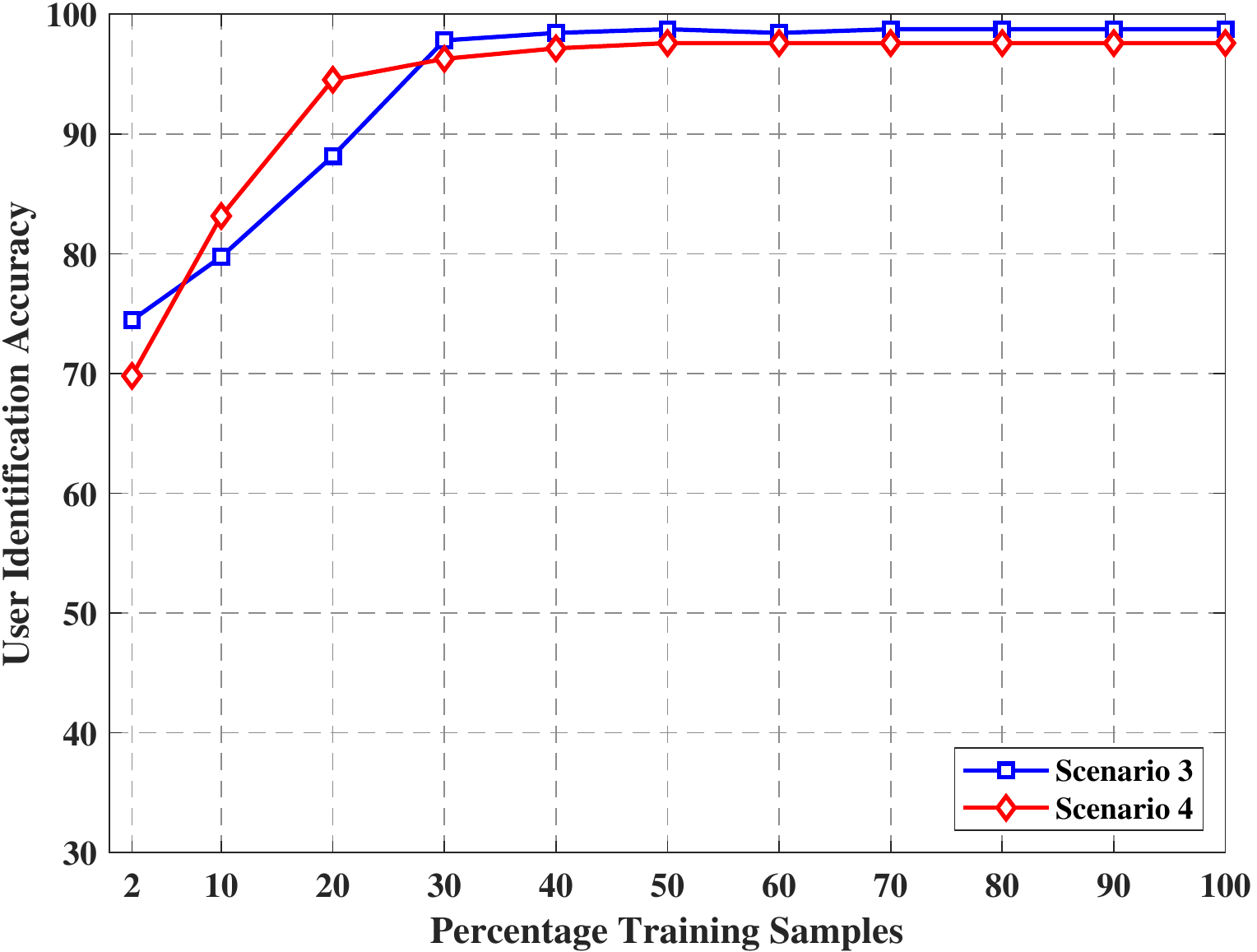}
	\caption{The figure presents a comparison of the user identification accuracy achieved by our proposed solution against varying dataset sizes in Scenarios $3$ and $4$. It is observed that for both scenarios $3$ and $4$, only $30\%$ of the training data is required to achieve the optimum performance.}
	\label{fig:percent_training_samples}
\end{figure}


\subsection{Can visual and wireless data be utilized for user identification?}\label{subsec:single_perf}

To answer this question, we evaluate the proposed single sample-based user identification solution on the development dataset of both scenarios $3$ and $4$ as described in \sref{sec:dev_data}. In particular, the proposed solution is trained and tested individually in both scenarios. In \figref{fig:combined_scene_score}, we plot the achieved user identification accuracy for both the scenarios. It is observed that for the input sequence length of $1$, the proposed solution achieved an accuracy of $98.43\%$ and $97.16\%$ for scenarios $3$ and $4$, respectively. The high accuracy of the proposed approach with just one observed sample highlights that sensing-aided solutions can enable user identification in a multi-candidate scenario. However, relying on just one sample to identify the user has its challenges. A key component of the proposed solution relies on the accurate detection of the objects of interest in the wireless environment. However, the state-of-the-art image-based object detection models have mean average precision (mAP) of $60 - 70\%$, which highlights that these models are imperfect. The inherent non-idealities of the real-world data further make it more challenging to detect objects accurately. This might result in the user itself not being detected, leading to errors in user identification. Another challenge arises when the user is blocked by other stationary and dynamic objects in the environment. By just relying on a single sample, it is not possible to detect the blocked user. A promising solution to overcome these limitations is to observe a sequence of image samples. It helps in  increasing the probability of object detection and, in general, improving user identification accuracy. Next, we present the performance of the proposed user-identification solution versus the percentage of labeled training samples required to achieve optimum performance. 


\subsection{How many training samples are needed for the user identification task?} \label{subsec:num_train_samples}
In the field of supervised machine learning, a critical factor is the availability of labeled datasets. However, obtaining these labels can often prove to be a formidable challenge. Given these inherent complexities, it becomes pivotal to determine the required number of training samples for the task of user identification. It aids in efficiently managing computational resources and navigating the challenges of labeled dataset procurement. \figref{fig:percent_training_samples} helps us delve into this issue, illustrating the relationship between user identification accuracies and the number of training samples employed in Scenarios $3$ and $4$. Achieving the optimum user identification accuracy as shown in \figref{fig:combined_scene_score} necessitates approximately $30\%$ of the total training samples. In particular, the proposed machine learning model is capable of learning the given user identification task with roughly $270$ samples for Scenario $3$ and approximately $310$ samples for Scenario $4$. In the next subsection, we present the performance of the proposed user-identification solution on the sequence data.

\begin{figure}[t]
	\centering
	\subfigure[]{\centering \includegraphics[width=1.0\linewidth]{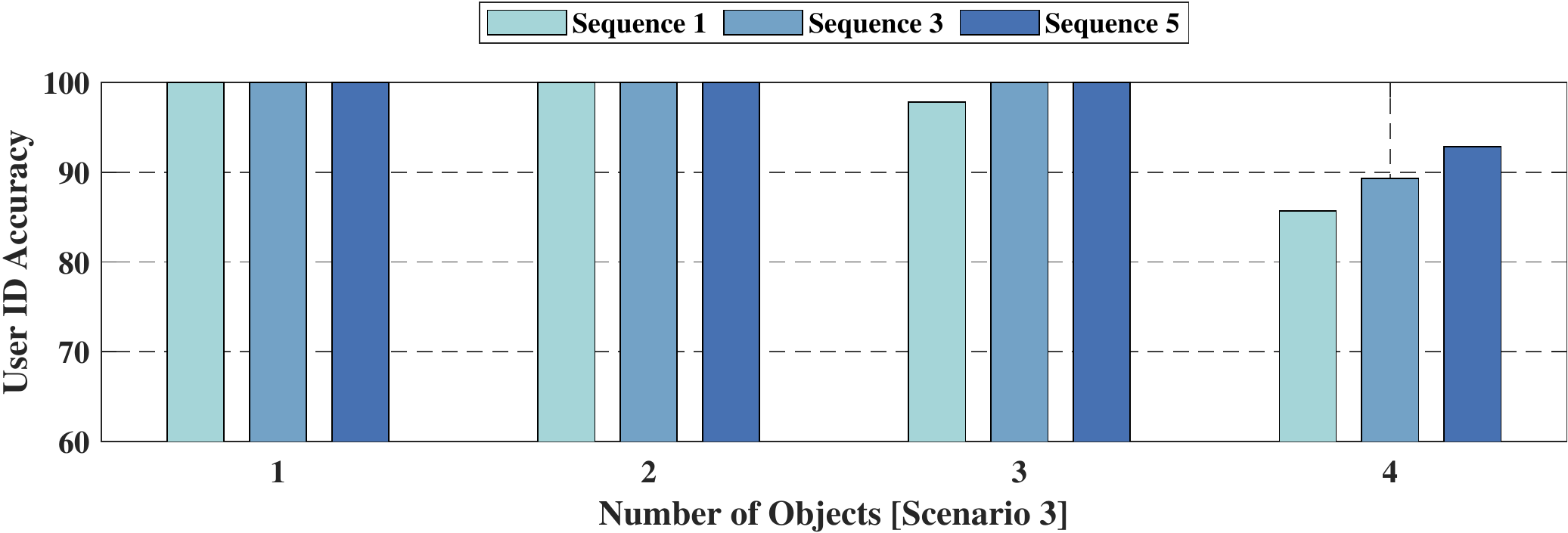}\label{fig:scenario3_num_obj_acc}}
	\subfigure[]{\centering \includegraphics[width=1.0\linewidth]{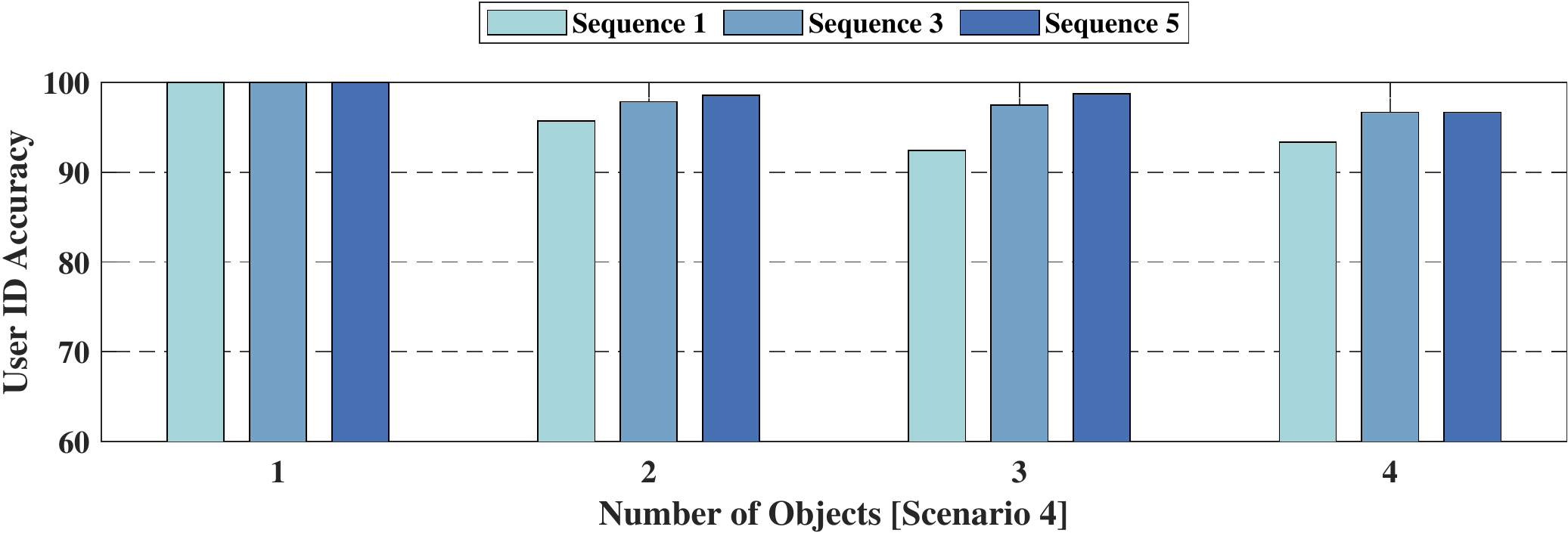}\label{fig:scenario4_num_obj_acc}}
	\caption{This figure presents the user identification accuracy versus the number of objects in the scene for both scenarios 3 and 4. It is observed that as the number of objects increase, the chances of mis-predictions increase. }
	\label{fig:num_obj_vs_acc}
\end{figure}


\begin{figure}[t]
	\centering
	\subfigure[]{\centering \includegraphics[width=1.0\linewidth]{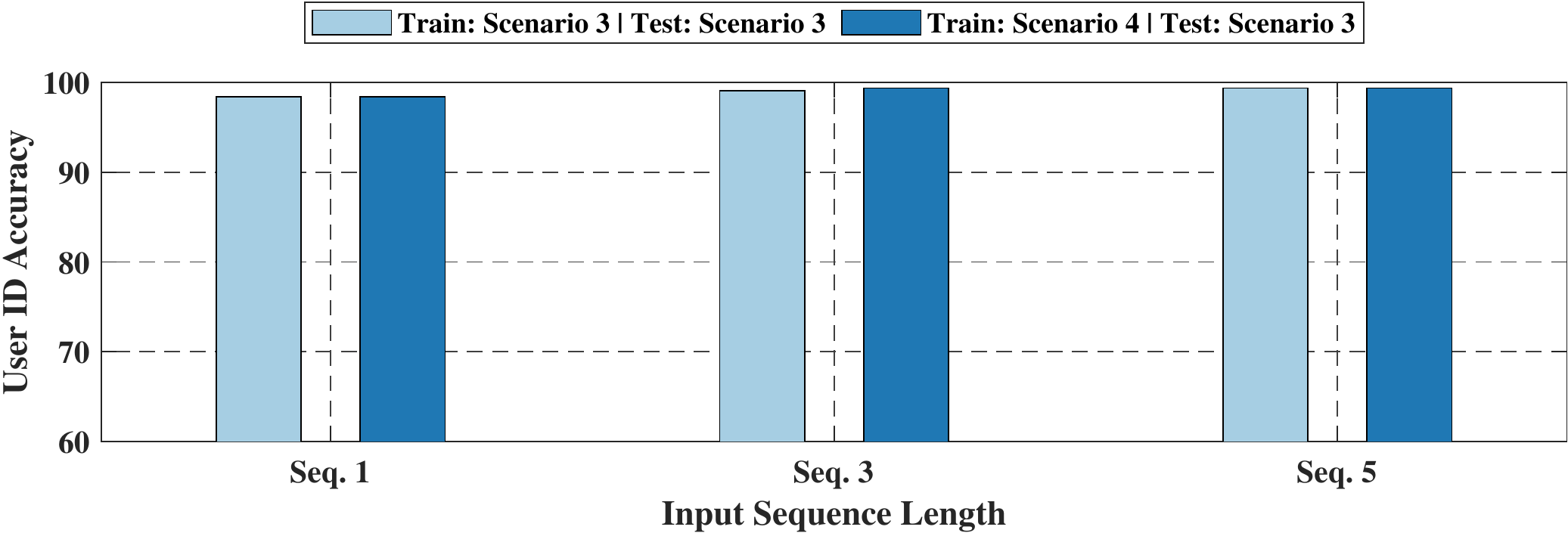}\label{fig:scenario3_4_acc}}
	\subfigure[]{\centering \includegraphics[width=1.0\linewidth]{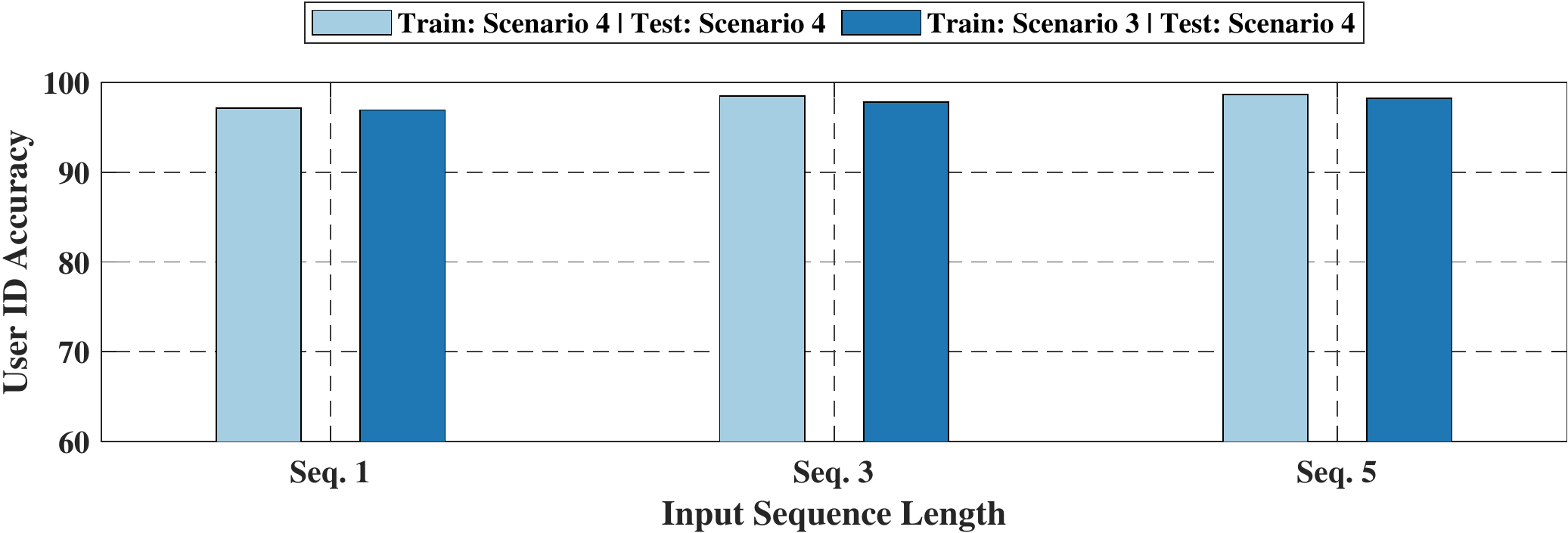}\label{fig:scenario4_3_acc}}
	\caption{This figure highlights the proposed solution's ability to generalize across different data distribution. The figures (a) and (b) presents the comparison between two cases: (i) The user identification accuracy of the proposed solution when trained and tested on the same scenario dataset. (ii) The model's performance when trained on one scenario training data and evaluated on the other scenario test data.  }
	\label{fig:cross_scenario_acc}
\end{figure}


\subsection{Does observing a sequence of data samples help?} \label{subsec:seq_perf}

In order to evaluate the effect of sequence data, we constructed a time-series dataset with a window length of $3$ and $5$ for both the scenarios $3$ and $4$. Furthermore, we extended the single sample-based solution to the sequence-based user identification solution as proposed in \sref{subsec:sequence_tx_id}. In Fig.~\ref{fig:combined_scene_score}, we present the user identification accuracy versus the input sequence length for both scenarios. It is observed that increasing the input sequence length for both scenarios helped achieve better identification performance. In order to further investigate the impact of observing a sequence of data samples, we plot the number of objects versus the prediction accuracy for both scenarios in Fig.~\ref{fig:num_obj_vs_acc}. It is important to highlight here that we have only plotted the performance up to $4$ objects. This is primarily because the number of sequences with more than $4$ objects is extremely small, as shown in Table~\ref{ref:tab_dataset_analysis}. Therefore, it is difficult to draw any meaningful conclusion from the performance of those sequences. In Fig.~\ref{fig:num_obj_vs_acc}, we observe an interesting trend; as the number of objects increases, the sequence-based approach achieves better performance. The improved performance can be attributed to the fact that as the number of objects in the scene increases, the chances of missed object detection, etc., also increase. Therefore, relying on just one sample will lead to reduced identification accuracy. The results, therefore, validate our initial intuition that observing a sequence of input data samples should perform better than just observing one data sample. 

\subsection{How accurate is the proposed distance-based object association step?}
In this section, we evaluate the effectiveness of our proposed distance-based tracking algorithm in the context of vehicle-to-infrastructure communication. The objective of this step is to assign and maintain object IDs within an image sequence, which is fundamental to our problem statement. To measure the accuracy of our object association step, we utilize the ground-truth bounding box of the user (transmitter) in the scene. In particular, we compare the IDs assigned to the user across all the samples in a sequence. A correct association is identified when the user has been assigned the same ID across all samples. The proposed solution achieves high object association accuracy across both Scenarios $3$ and $4$, irrespective of the sequence length. Specifically, we achieve an accuracy of $100\%$ for a sequence length of $3$. For the sequence length of 5, the accuracy remains robust, slightly reducing but still close to $\approx 99.50\%$. One possible explanation for such high accuracy is the predictable motion paths of vehicles in V2I scenarios, which makes the proposed Euclidean distance-based object tracking algorithm particularly effective. This consistent performance demonstrates our approach's adaptability and effectiveness in identifying and tracking objects throughout various image sequences.

\subsection{Does variation in data distribution impact the model's performance?} \label{subsec:cross_scenario_analysis}

In \sref{subsec:single_perf} and \sref{subsec:seq_perf}, we presented the user identification performance of the sensing-aided solution for single-sample and sequence-based approaches, respectively. It is observed that for these approaches, the proposed solution can identify the user with high fidelity. However, the proposed solution was trained and evaluated on the same scenario dataset in these experiments. Although such an experimental design is necessary to develop initial insights and understanding, it is insufficient. In order to develop solutions that can eventually be deployed in the real world, we need to test their capability further. One such test is the model's ability to generalize across different data distributions. Any shift in the data distribution, i.e., the distribution of the data differs between the training and test stages, can adversely impact the model's performance. For example, if a model is trained on images collected only during the day, its performance can drop if tested at night. Therefore to study the proposed solution's ability to generalize, we design an inter-scenario experiment. Here, we train the model on the training dataset of one scenario and evaluate the performance on the test dataset of another scenario. In particular, we utilize the same scenarios, $3$ and $4$, for this experiment. Although scenarios $3$ and $4$ belong to the same location, the data samples were collected during different times of the day. In Figure~\ref{fig:cross_scenario_acc}, we present the user identification performance achieved by the proposed solution for this inter-scenario experiment. More specifically, we compare the performance between the two cases: (i) train and test on the same dataset versus (ii) train on one scenario and evaluate on a different scenario test data. In Fig. \ref{fig:scenario4_3_acc} see a slight drop in the user-identification accuracy when the model trained on scenario $3$ training dataset is evaluated on the scenario $4$ test dataset. In general, we observe a $1-2\%$ drop in accuracy for this case. Scenario $4$ being collected during the night time, it is generally more challenging for object detection and, hence,the downstream user identification task. Overall, from Figures~\ref{fig:scenario3_4_acc}, \ref{fig:scenario4_3_acc}, we observe that the proposed solution can efficiently generalize across the two scenarios.

\begin{figure}[!t]
	\centering
	\includegraphics[width=1.0\linewidth]{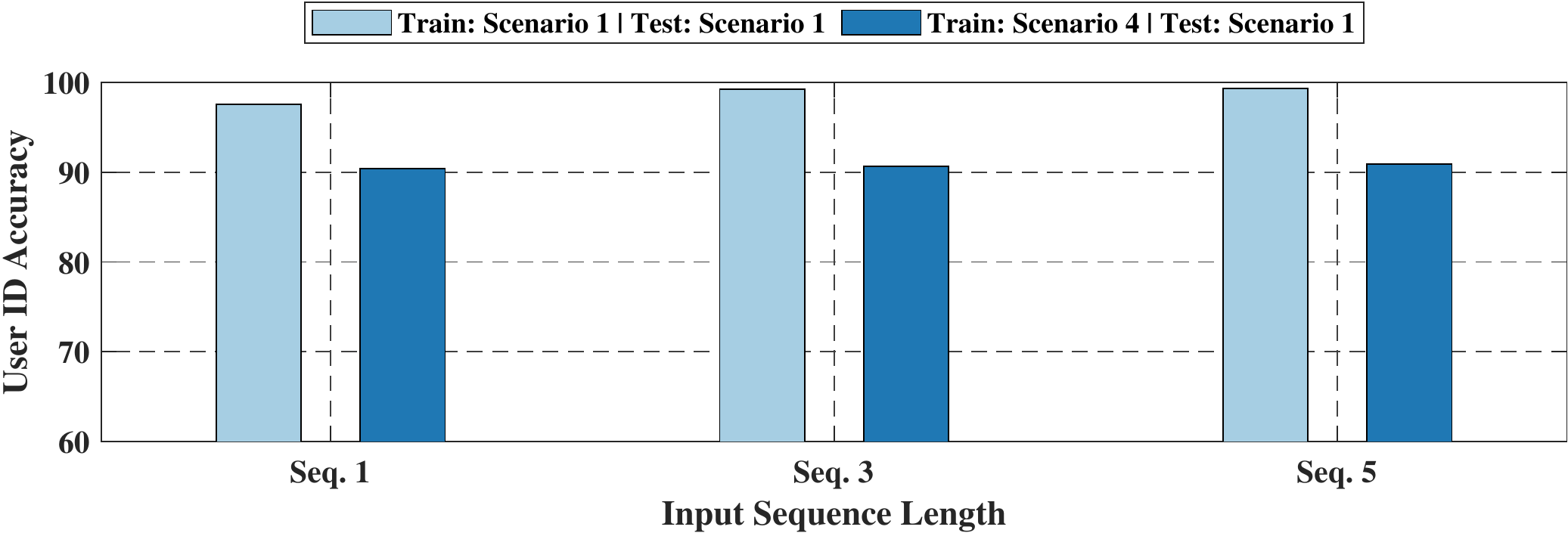}
	\caption{This figure shows the proposed solution's ability to adapt to unseen scenarios. We utilize scenario $1$ and $4$ dataset for this experiment. Here, we present the comparison between two cases: (i)The user identification accuracy of the proposed solution when trained and tested on scenario $1$ dataset alone. (ii) The model's performance when trained on scenario $4$ training data and evaluated on scenario $1$ test data.  }
	\label{fig:scenario1_4_adaptation}
\end{figure}

\begin{figure}[t]
	\centering
	\subfigure[Scenario 3]{\centering \includegraphics[width=1.0\linewidth]{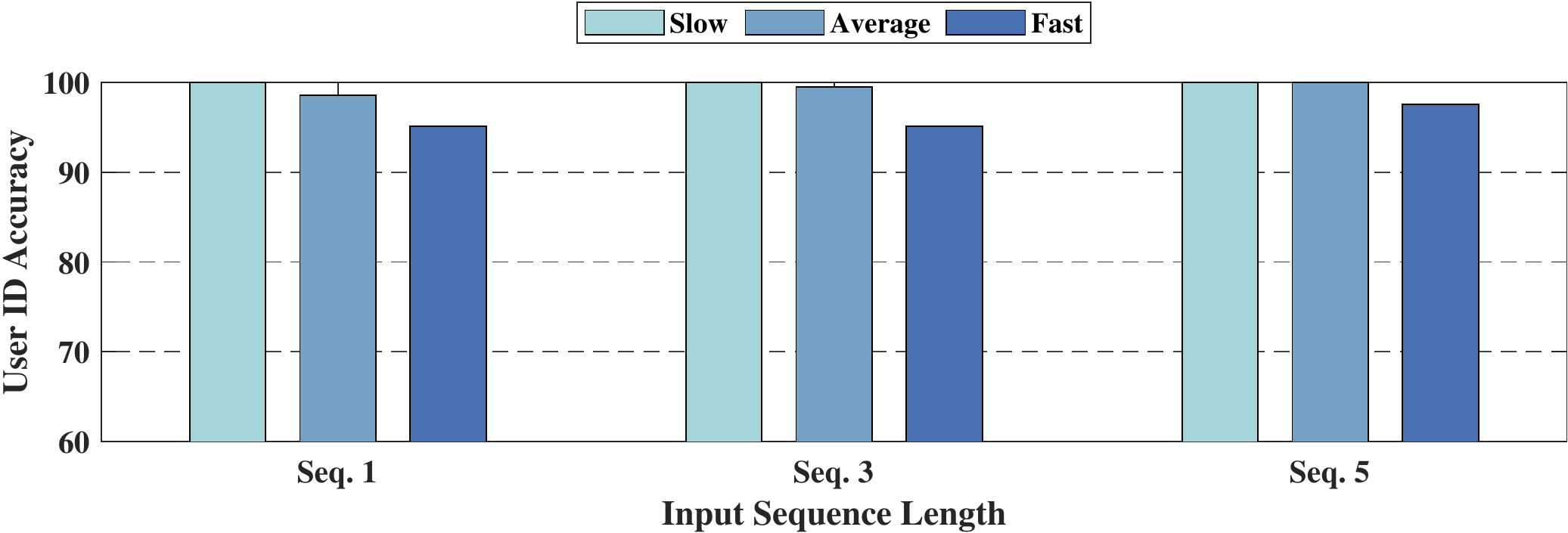}\label{fig:scenario3_speed_acc}}
	\subfigure[Scenario 4]{\centering \includegraphics[width=1.0\linewidth]{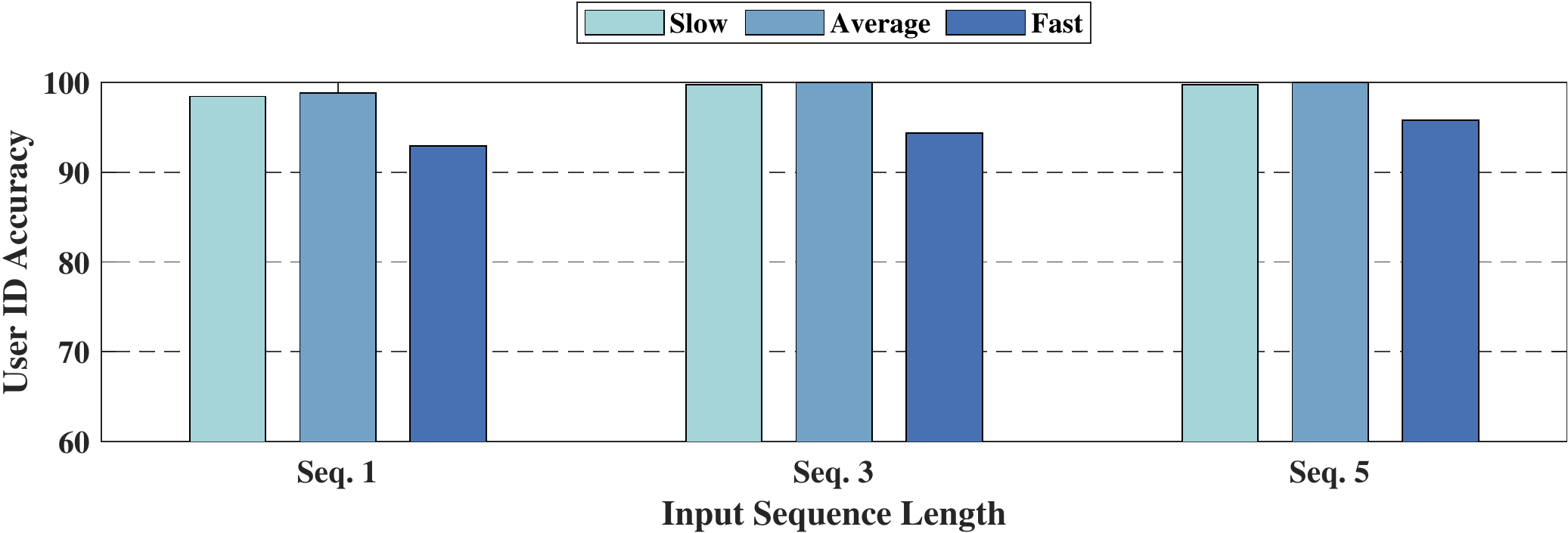}\label{fig:scenario4_speed_acc}}
	\caption{The figure shows the impact of the user speed on the user-identification performance of the proposed solution. In (a) and (b), the user identification accuracies versus the user speed is presented for scenarios $3$ and $4$, respectively. }
	\label{fig:speed_vs_accuracy}
\end{figure}



\subsection{Can the proposed solution adapt to unseen scenarios?} \label{subsec:adaptation}

In \sref{subsec:cross_scenario_analysis}, we investigate the proposed solution's ability to generalize across different data distributions. Another challenge towards real-world deployment is the fast adaptation to unseen scenarios. Given that the $5$G and beyond basestations will be deployed across different locations developing site-specific machine learning models, i.e., different models trained in a supervised fashion for each location, is not feasible. This is primarily due to the unavailability of such labeled datasets for each location. Overcoming this challenge necessitates the development of efficient solutions that can adapt quickly to an unseen location with few or no labeled data samples. To evaluate such adaptation capability of the proposed solution, we utilize scenarios $1$ and $4$ of the DeepSense $6$G dataset. These two scenarios were collected at different locations and at different times of the day. As shown in Figure~\ref{fig:deepsense_scenario_photos}, scenario $1$ consists of a $2$-lane street, whereas scenario $4$ is a $6$-lane street.  Further, the distance of the basestation from the street differs for these two locations. Such differences result in variations in the distribution of the mmWave receive power. All these make it highly challenging for any solution to adapt to an unseen scenario. In Figure~\ref{fig:scenario1_4_adaptation}, we present the achieved user identification accuracy for two cases: (i) The proposed ML model is trained and tested on scenario $1$ dataset alone, and (ii) The model has trained on scenario $4$ training dataset and evaluated on the test set of scenario $1$. We observe a $7-9\%$ drop in accuracy between the two cases, signifying how challenging this adaptation task is. The key takeaway is that even with no training data from scenario $1$, the proposed solution can identify the users with $\approx 90\%$ accuracy. Such a performance highlights the ability of the proposed solution to adapt to unseen scenarios.


\subsection{Does user-speed impact the user identification performance?} \label{subsec:user_speed_analsysis}
Given the dynamic nature of the dataset, the speed of each user (vehicle) varies with time. Therefore, it is important to consider the impact of vehicle speed on the user identification accuracy. In  order to calculate the user speed, we utilize the position of the user (available as part of the DeepSense 6G dataset). In particular, we estimate a user speed by considering the difference between the initial and final position in each sequence with $5$ samples and divide it by $5$. Further, we calculate the speed mean $\tilde{\mu}$ and standard-deviation $\tilde{\sigma}$. Using $\tilde{\mu}$ and $\tilde{\sigma}$, we divide those users into three buckets: (i) slow-moving user with speeds less than or equal to $\tilde{\mu} - \tilde{\sigma}/2$; (ii) fast-moving user with speeds greater than or equal to $\tilde{\mu} + \tilde{\sigma}/2$; and (iii) average-speed user with speeds between those of slow- and fast moving users. In \figref{fig:speed_vs_accuracy}, the user identification accuracy versus the vehicle speed is presented. It is observed that for both scenarios $3$ and $4$, the slower moving users result in better user identification accuracy. However, in most of the samples, the difference in accuracy between the slow and fast moving user is very small. This further highlights the model's ability to identify the user even for fast-moving vehicles with very high confidence.

\section{Conclusion}\label{sec:conc}

This paper explores the potential of leveraging visual and mmWave wireless data for identifying the probable user in the wireless environment. It takes an essential step toward addressing the concern about the practicality of the sensing-aided wireless communication solution in real multi-object communication settings. It does so by (i) defining the novel user identification task, (ii) proposing a deep learning-based solution, and (iii) extending the solution from a single sample-based approach to a sequence-based solution. The key takeaways of evaluating our proposed user identification solution based on the large-scale real-world dataset, DeepSense 6G, can be summarized as follows: (i) Even with a single data sample, the proposed solution achieves $\approx 97\%$ user identification accuracy. (ii) For data samples with more objects in the wireless environment, observing a sequence of previous data samples helps achieve better prediction accuracy than just the current data sample. These results highlight the potential gains of leveraging visual and wireless data in identifying probable users in the wireless environment and place more emphasis on designing better algorithms to tap into the wealth of information in the input sensing data.

\balance

\end{document}